\documentclass[10pt,a4paper]{article}

\usepackage[utf8]{inputenc}
\usepackage{amsmath}
\usepackage{amsfonts}
\usepackage{amssymb}
\usepackage{amsthm,epsfig,epstopdf,titling,url,array,float}
\usepackage{multirow}
\usepackage{multicol}
\usepackage{geometry}
\usepackage{rotating}
\usepackage{setspace}
\theoremstyle{plain}
\newtheorem{thm}{Theorem}[section]
\newtheorem{lem}[thm]{Lemma}

\theoremstyle{definition}

\theoremstyle{remark}

\usepackage{mathtools}
\newcommand\givenbase[1][]{\:#1\lvert\:}
\let\given\givenbase

\DeclarePairedDelimiterX\Basics[1](){\let\given\sgiven #1}

\begin{document}

\title{Conditional Masking to Numerical Data}

\author{Debolina Ghatak and Bimal K. Roy}
\date{}
\maketitle

\begin{abstract}
Protecting the privacy of data-sets has become hugely important these days. Many real-life data-sets like income data, medical data need to be secured before making it public. However, security comes at the cost of losing some useful statistical information about the data-set. Data obfuscation deals with this problem of masking a data-set in such a way that the utility of the data is maximized while minimizing the risk of the disclosure of sensitive information. Two popular approaches to data obfuscation for numerical data involves (i) data swapping and (ii) adding noise to data. While the former masks well sacrificing the whole of correlation information, the latter gives estimates for most of the popular statistics like mean, variance, quantiles, correlation but fails to give an unbiased estimate of the distribution curve of the original data. In this paper, we propose a mixed method of obfuscation combining the above two approaches and discuss how the proposed method succeeds in giving an unbiased estimation of the distribution curve while giving reliable estimates of the other well-known statistics like moments, correlation. 
\end{abstract}


\section{Introduction}
Estimation of statistics like mean, variance, quantiles is fundamental for statistical analysis of data. But, if the data is sensitive to the individual bearing the information, it may be almost impossible to publish the data in its raw form. Data obfuscation calls for methods that can protect the individual information from any possible intruder while retaining as much statistical utility as possible. The two motives of data obfuscation are 
\begin{enumerate}
\item[(i)] Maximising data utility
\item[(ii)] Minimising risk of disclosure.
\end{enumerate} 

A typical data-set we are thinking of consists of $m$ variable values corresponding to $n$ individuals and among these $m$ variables one or a few variables have some sensitive information related to the individual. One may think, at first, that if the name or identification number of the individual is erased, there is no point in protecting the values as the intruder would not know who this value belongs to. But, in reality, the scenario is different from this intuitive belief. In many cases, even if the identification information is not given, looking at all the variable values in the row, the intruder often becomes successful in identifying the individual. The paper \cite{TSBM} discusses how the presence of a few non-sensitive attributes may jointly disclose the identity of an individual in a data-set consisting of several attributes. For example, in an official data-set, age-group, sex, birthplace may not individually leak the identity of an individual but suppose, it is known to the intruder that there is only one male in the data-set with age group 20-25 who comes from ``Kerala". Then jointly, these three attribute values will reveal the identity of the individual. Thus, practically, it is often very difficult to hide the identity of the individual in presence of many attributes. Moreover, in many practical cases, the identifier may be an essential attribute. Thus, the sensitive attributes, i.e., the ones that carry values that shall not be disclosed as for example income data, marksheet data etc. needs protection. 
\medskip

\noindent
As an early reference one can go through the papers of Steinberg and  Pritzker ( 1967) \cite{JSLP} Bachi and Banon ( 1969) \cite{RBRB} that discusses the importance of privacy protection in the case of sensitive data. Dalenius ( 1974)\cite{TD} discussed the importance of the disclosure problem in statistical studies. Mugge (1983) \cite{RHM} discussed the issues in protecting confidentiality in National Health Statistics. Dalenius ( 1977a)\cite{TDA} Fienberg ( 1994)\cite{SEF} discusses the importance of the problem in Computers and  individual privacy.
\paragraph*{}
There are various methods of obfuscating data as discussed in \cite{WAF}\cite{WF}\cite{MRA}\cite{DR}. Two of them involve swapping the data values among them and adding noise to individual values. While in a large data-set if values corressponding to only one sensitive variable are swapped, it does not harm the estimates of mean, variance, moments or quantiles for the data but, its correlation information with any other variable is completely erased after swapping. This gave rise to methods like rank swapping as discussed by Moore( 1996) in \cite{MRA}, and data shuffling by R. Sarathy and K. Muralidhar in \cite{RK}. To get satisfactory estimates after rank swapping, the disclosure risk gets very high. Hence, this method often fails to perform well. Data shuffling is a more reliable method of obfuscation; it shuffles the data in such a way that the correlation between variables is not harmed, but in case there are a large number of variables it may become impossible or much more tedious to shuffle even the non-sensitive variables to retain their correlation information with the sensitive one.

Adding noise to data, on the other hand, gives us estimates of the mean, variance and quantiles without doing much harm to the correlation information. But after sufficient obfuscation, there may be a considerable amount of loss of utility of the data. Also, there is no known procedure to get an unbiased estimation of the distribution curve in this model which makes quantile estimation, a hard problem. 

In this paper, we introduce a new approach to data obfuscation, a method that combines data swapping and addition of noise. Here, a part of the data is swapped and to the rest of the data, we add noise from a normal distribution with mean zero and known variance. The resulting data in hand will be useful in giving an unbiased estimation procedure to the distribution curve of the original data resulting in very good quantile estimates while giving sufficient masking to the data values.

In Section 2, we discuss the basic model of the procedure, estimate the disclosure risk and also some useful statistics with the required proofs in the Appendix section. In Section 3, we simulate a data-set of size 2000 and apply our procedure to the given problem to see check how our process works. Also, a simulation study is given to see how the process works for increasing sample size. In Section 4, we apply the procedure to a real-life problem, a data-set of marks of 445 students of an institute and see how after sufficient obfuscation our procedure gives reliable estimates of various statistics, especially the quantiles. Finally, we conclude with some discussions in Section 5.

\section{Basic Problem}

Suppose we have a large data-set with $m$ variables, corresponding to $n$ individuals; and among these $m$ variables, there is some numerical variable which is sensitive and needs to be protected. Let the data values corresponding to this variable be $\{X_i, 1\leq i\leq n\}$, which is assumed to come from some unknown distribution function $\{G(x),x \in R\}$ which is continuous. The idea is to obfuscate the data in a way such that with some probability $p$ the obfuscated value $Z_i$ corresponding to $X_i$ takes any other value among $\{X_i, 1\leq i\leq n\}$ expect $X_i$ and with probability $1-p$ it adds a noise to the data values. To perform this method, we first simulate $\{B_i, 1\leq i\leq n\}$ independent of $\{X_i, 1\leq i\leq n\}$ where $B_i$ is a $Bin(1,p)$ variable, i.e, a binary variable with probability of success $p$. The obfuscated data $\{Z_i, 1\leq i\leq n\}$ then looks like the following

\begin{equation}
   Z_i = \left\lbrace
     \begin{array}{lr}
       X_j & \mbox{   $j\neq i$,  if $B_i=1$ } \\
        X_i+Y_i & \mbox{if $B_i=0$}
     \end{array}
   \right.
   \label{Model:CM}
\end{equation}

where $Y_i \sim N(0,\sigma^2)$, $\sigma$ is known, $1 \leq i \leq n$, and $j$ is chosen randomly from the set $\{\{1,2,\ldots,n\} \backslash \{i\}\}$ randomly, i.e., with probability $\frac{1}{n-1}$.

Note that $\{Z_i, 1\leq i\leq n\}$ is a set of independent variables, each data point $Z_i$ being dependent on $X_i$ and $B_i$ but not on each other. However, two $Z_i$s may take the same value but usually with a very low probability for large value of $n$. Hereafter, we will discuss the estimation of the raw moments( mean and variance especially) and quantiles of $X$ from the knowledge of $\{Z_i, 1\leq i\leq n\}$, $p$ and $\sigma$, its correlation with any other variable $X^{\prime}$ and also about the disclosure risk associated with such model. Usually, Laplace noise is used in additive noise model but here we use Normal error because under this model, estimation would become very hard for any other distribution of noise other than Normal.

\subsection{Estimation of Moments}

A fundamental problem in statistics is to estimate the moments of a data set especially the mean and variance. Here, we will see that even after sufficient obfuscation it is possible to get reliable estimates for the raw moments of the true data using the following theorem. Once we find the raw moments, central moments can be easily derived using the standard relation between moments.

\begin{thm}{
If $\{X_i, 1\leq i\leq n\}$ is assumed to be an i.i.d. sample from some unknown distribution function $G(x)$} ( $G$ is a continuous function) with finite absolute raw moments, i.e., $$E(|X_i|^k)< \infty \mbox{ , $\forall$ $k \in N$}.$$ and $\{Z_i, 1\leq i\leq n\}$ is obtained using Equation~\eqref{Model:CM}, then an unbiased estimator for the $k^{th}$ raw moment of $X$ ( $X \sim G(.)$, $k \in N$) is obtained from the recursion relation given below
$$\hat{\mu}_{(X,k)}=\hat{\mu}_{(Z,k)}-(1-p)\cdot ({\mu}_{(Y,k)}+\binom{k}{1}\cdot {\mu}_{(Y,k-1)} \cdot \hat{\mu}_{(X,1)}+ \cdots + \binom{k}{k-1}\cdot\hat{\mu}_{(X,k-1)}\cdot {\mu}_{(Y,1)})$$
where, $\hat{\mu}_{(X,1)}=\bar{Z}$, $\hat{\mu}_{(Z,k)}=\frac{1}{n}\cdot \sum_{j=1}^n{Z_j^k}$, 
$$ \begin{array}{lr}
{\mu}_{(Y,k)} &=\mbox{$k^{th}$ raw moment of $N(0,\sigma^2)$} \\
&=\left\lbrace
     \begin{array}{lr}
       0 & \mbox{  $k$ is odd } \\
       {\frac{\sigma^k \cdot k! }{2^{k/2} \cdot {k/2}!}}  & \mbox{  $k$ is even.}
     \end{array}
   \right.
\end{array}$$
\end{thm}

From Theorem 2.1, we have $\bar{Z}$ to be an unbiased estimate of $\mu_X$, mean of $X$.

Also, we have,

$$ \begin{array}{lr}
\hat{\mu}_{(X,2)} &= \hat{\mu}_{(Z,2)} - (1-p)\cdot ({\mu}_{(Y,2)}+2\cdot {\mu}_{(Y,1)} \cdot \mu_X) \\
&= \hat{\mu}_{(Z,2)} - (1-p)\cdot \sigma^2
\end{array}$$

Define, $\hat{S}_X^2 = \hat{S}_Z^2 -(1-p)\cdot \sigma^2$ where ${\hat{S}_Z^2 =\frac{1}{n-1}\sum_{i=1}^n{(Z_i - \bar{Z})^2}}$.
$$ E(\hat{S}_X^2) = \sigma_Z^2 - (1-p)\cdot \sigma^2 
= \sigma_X^2 \mbox{ , where ${\sigma_X^2=Var(X_1),\sigma_Z^2=Var(Z_1)}$}$$

\subsection{Estimation of Quantiles}

In non-parametric studies, the estimation of the median, or in general, any quantile is most crucial. Moreover, quantiles are robust statistics and hence is not much affected by the presence of outlying data-points. To do the same we first estimate the c.d.f. of $X$ from the model using the following Theorem.

\begin{thm}
If $\{X_i, 1\leq i\leq n\}$ is assumed to come from some unknown d.f. $\{G(x), x \in R\}$ and $\{Z_i, 1\leq i\leq n\}$ is obtained using Equation~\eqref{Model:CM} then, for $p>0.5$,
\begin{equation}
 T_1(x)=\frac{1}{np}\sum_{j=1}^n{\sum_{t=0}^\infty{\lambda^t\cdot \Phi_{\sigma\sqrt{t}}(x-Z_j)}}
 \label{T:1}
\end{equation}
is an unbiased estimator for $G(x)$ $\forall x \in R$ , where $\lambda=-\frac{1-p}{p}$ and $\Phi_m(i)$ is the cumulative distribution function of a Normal variable at $i$ with mean 0 and standard deviation $m$ for $m>0$ and for $m=0$, $\Phi_0(i)=\mathbb{I}(x-Z_j)$ where $\mathbb{I}(i)=1$ if $i \geq0$ and $0$ o.w.
\end{thm}

Although Theorem 2.2 gives us an u.e. for $G(x)$, this estimator will give good results only for very large $n$. If $n$ is large but not very large, then sometimes the estimator we get to look at in the following Theorem may give us a smooth estimate to the curve of $G(x)$, i.e., $\hat{G}(x)$ is a smooth function or a function that has derivatives of all orders everywhere in its domain.

\begin{thm}
If $\{X_i, 1\leq i\leq n\}$ is assumed to come from some unknown distribution function $\{G(x), x \in R\}$, having a density function, and $\{Z_i, 1\leq i\leq n\}$ is obtained using Equation~\eqref{Model:CM} then, for $p>0.5$,
\begin{equation}
 T_b(x)=\frac{1}{np}\sum_{j=1}^n{\sum_{t=0}^\infty{\lambda^t\cdot \Phi_{b_t}(x-Z_j)}}
 \label{T:b}
\end{equation}
is a smooth estimator for $G(x)$ $\forall x \in R$ , where $\lambda=-\frac{1-p}{p}$, $b_t=\sqrt{tb^2+\sigma^2}$ and $\Phi_\upsilon(i)$ is the cumulative distribution function of a Normal variable at $i$ with mean 0 and standard deviation $\upsilon$.
\end{thm}


The estimator $T_1(x)$ is unbiased while $T_b(x)$ is not but is smooth and sometimes more useful than the former. However, we study the asymptotic properties of such estimators in the following theorem.

\begin{thm}
Both the estimators $T_1(x)$ and $T_b(x)$ are the mean of i.i.d. random variables with finite expectation and variances and are consistent in the sense
$$ T_i(x) \overset{P}{\longrightarrow} G(x) \mbox{ , $\forall x$ as $n \rightarrow \infty$ , $i=1,b$} $$
\end{thm}
To estimate the $\alpha^{th}$ quantile, we equate $\hat{G}(x)$ instead of $G(x)$ with $\alpha \in (0,1)$ using some numerical method, i.e., $\hat{\xi_\alpha}$ is such that 
$$ \hat{G}(\hat{\xi_\alpha})=\alpha $$

\subsection{Estimation of correlation}

While swapping of data values among them retains the exact information of the mean, variance and quantiles, it does completely erase any possible correlation information of the variable with any other variable. That is why additive noise model is used while obfuscating a single sensitive variable among many. However, our method gives us a reasonable estimate for the correlation coefficient.

\begin{thm}
Let $\{X^{\prime}_i, 1\leq i\leq n\}$ be the data points corresponding to some variable $X^{\prime}$ associated with $X$, correlation coefficient between $X$ and $X^{\prime}$ being $\rho_{(X,X^{\prime})}$. Then, if the variance of $X$ and $X^{\prime}$ exists and is finite and so is $d(X,X^{\prime})=E[(X-E(X))^2(X^{\prime}-E(X^{\prime}))^2]$,
$$ \hat{\rho}_{(X,X^{\prime})} = \frac{1}{1-p} \cdot \frac{\hat{Cov}(Z,X^{\prime})}{\sqrt{\hat{Var}(X^{\prime})}\cdot \sqrt{\hat{Var}(X)}} $$
is a consistent estimator of $\rho$ where $\hat{Cov}(Z,X^{\prime})=\frac{1}{n}\cdot[\sum_{j=1}^n{Z_j\cdot X^{\prime}_j}-\bar{Z}\cdot \bar{X^{\prime}}]$.
\end{thm}
\subsection{Disclosure Risk}

In data obfuscation, estimation is important but not at the cost of disclosure of data values. So to check the disclosure risk of data, obfuscated with this method, note that the possible estimators for $X_i$ from the available information of obfuscated data are $Z_i$ and $\bar{Z}$, both being unbiased estimators of $X_i$. $\bar{Z}$ can be shown to be a better estimator of $X_i$ than $Z_i$, as long as the minimum MSE is concerned, which means we get one estimate for each $X_i$ which is possibly its best estimator. Hence, the disclosure risk is not expected to be high for medium to large values of $n$.
\begin{eqnarray*}
MSE(Z_i)=E(Z_i-X_i)^2 & = & \frac{p}{n-1}\cdot \sum_{j=1}^n{E(X_i-X_j)^2} + (1-p)\cdot E(Y_i^2) \\
& = & 2p\cdot \frac{n}{n-1} \sigma_X^2 +(1-p)\cdot \sigma^2
\end{eqnarray*}

$\begin{array}{ll}
MSE(\bar{Z}) &= E(\bar{Z}-X_i)^2 \\
&= E((\bar{Z}-E(X))-(X_i-E(X)))^2 \\
&=E(\bar{Z}-E(X))^2+E(X_i-E(X))^2 -2E(\bar{Z}-E(X))(X_i-E(X)) \\
&=\sigma_X^2 +\frac{p\sigma_X^2 + (1-p)\sigma^2}{n} - 2\frac{\sigma_X^2}{n} \\
&=\sigma_X^2(1-\frac{2-p}{n}) + \frac{1-p}{n}\sigma^2 \\
&\leq MSE(Z_i) \mbox{ , since $2p\frac{n}{n-1}>1$ and $0 \leq \frac{2-p}{n} \leq 1$ for $n>1$.}
\end{array}$
 
 \paragraph*{}
 Thus, $\bar{Z}$ is a better estimator of $X_i$ than $Z_i$ for all $n>1$ as far as MSE is concerned. However, the disclosure risk for any estimator $\tau$ for $X_i$ can be measured by,
 $$ P[|\tau-X_i|<d] \mbox{ , for $d>0$}$$
i.e., the probability that $X_i$ lies within a $d$-boundary of its estimator. This measure was used in our previous work as discussed in \cite{GR}. For $S$ simulations, an estimate of risk is given by,
$$\frac{\sum_{s=1}^S{I_{[\tau_s \in (X_i-d,X_i+d)]}}}{S}$$

where $\tau_s$ is the estimate of $X_i$ for $s^{th}$ simulation and $I_{[A]}=1$ if event $A$ occurs and $0$ otherwise.

\section{Simulation Results}

To apply the discussed method we simulate a sample of dimension $n \times 2$ for a given copula structure using the function \emph{mvdc} in package \emph{copula} of \emph{R 3.3.2}. We take $n=2000$ and a copula structure such that the first variable is from \emph{Laplace($\mu_1=10$,$\sigma_1=1000$)}, the second variable from \emph{Laplace($\mu_2=50$,$\sigma_2=250$)} and the correlation between these two variables is $\rho=-0.7$. Of the two samples of size $n$, we consider the first one to be the sensitive variable and the other one the variable correlated with the sensitive one. Now we obfuscate the sensitive variable, say, $\{X_{i}, 1\leq i\leq n\}$ with the method discussed (Equation~\eqref{Model:CM}) to get the obfuscated variable $\{Z_{i}, 1\leq i\leq n\}$; we used $\sigma=\sigma_1$. Also, to compare the method with the general additive noise model, we also obfuscate $\{X_{i}, 1\leq i\leq n\}$ with addition of noise $\{Y_{1i}, 1\leq i\leq n\}$, $Y_{1i} \sim Laplace(0,\tilde{\sigma}^2)$ to get $\{Z_{1i}, 1\leq i\leq n\}$. Note that, the noise is taken from Laplace because we need to estimate the quantiles from the obfuscated data and as discussed in our previous work \cite{GR}, Laplace is till now the best possible choice for the obfuscating distribution. To keep the dispersion of the variables more or less same, we take $\tilde{\sigma}=\sigma_1$, although this is just a choice; in general, the method works well when the estimation of quantiles is needed from a large data-set with low disclosure risk. The method of obfuscation also works well for lower values of $\sigma$. 

\begin{figure}
\centering
\includegraphics[scale=0.4]{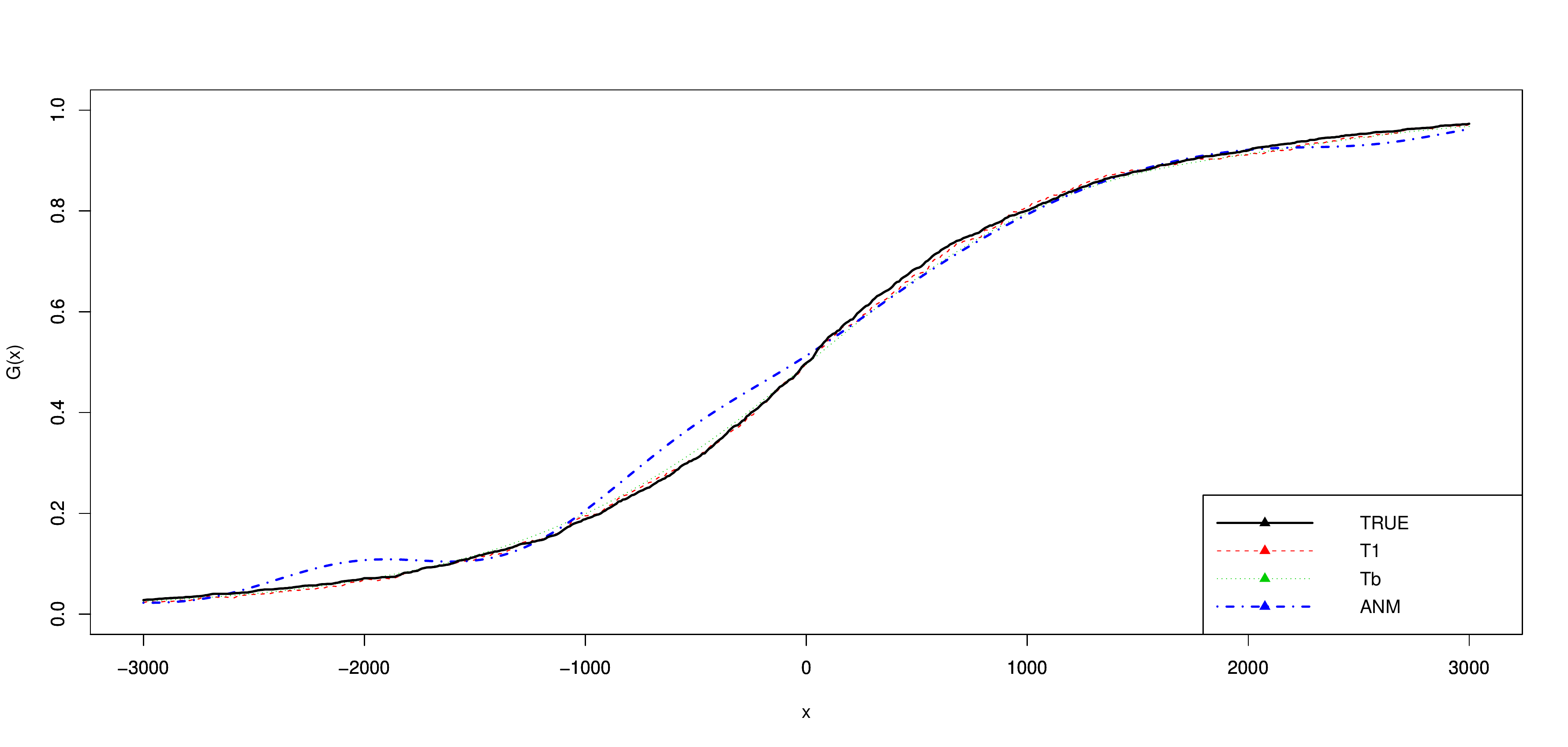}
\caption{{True and estimated distribution curve using Additive Noise Model and our method for simulated data}}
\label{graph:cond_mask}
\end{figure}

Figure~\ref{graph:cond_mask} shows the true distribution curve $G(x)$ and estimates of ${G(x)}$ from the obfuscated data-sets as discussed in the previous paragraph. Table~\ref{Quantiles:200} shows the true and estimated values of the quantiles for $\alpha=\{0.1,0.2,0.3,0.4,0.5,0.6,0.7,0.8,0.9\}$ and also that of the mean, variance and correlation with the other variable for the same data-set. Although theoretically we only have $p>0.5$, a very high value of $p$ tends to make the procedure a data swapping process which completely erases the correlation information. Thus, throughout this section, we will take $p=0.6$.

\begin{table}
\begin{center}
\caption{Estimated Statistics from original and obfuscated data using additive noise model and our method  for simulated data.}
{\begin{tabular}{|l|c|c|c|c| } 
\hline
Statistic & TRUE & T1 & Tb & ANM \\
\hline
0.1 & -1599.438 & -1720.259 & -1768.946 & -1818.652 \\
0.2 & -906.291 & -942.292 & -983.028 & -1147.267 \\
0.3 & -500.826 & -479.118 & -545.852 & -619.269 \\
0.4 & -213.144 & -191.17 & -217.001 & -255.806 \\
0.5 & 10 & 71.045 & 53.987 & 37.518 \\
0.6 & 233.144 & 296.573 & 313.607 & 337.252 \\
0.7 & 520.826 & 544.369 & 600.41 & 679.548 \\
0.8 & 926.291 & 924.456 & 976.201 & 1123.431 \\
0.9 & 1619.438 & 1631.023 & 1631.599 & 1814.377 \\
Mean & 10 & 10.27 & 10.27 & 10.834 \\
s.d. & 1414.214 & 1442.762 & 1442.762 & 1444.929 \\
Cor & -0.7 & -0.745 & -0.745 & -0.69 \\ 
\hline
\end{tabular}}
\label{Quantiles:200}
\end{center}
\end{table}

Looking at the graph, one can easily notice that the curves we get using $T_1$ and $T_b$ are comparatively closer to the cumulative distribution functions of $X$ than the one we get using the Additive Noise Model. Also, from the table we see the quantile estimates are improved from the Additive Noise Model without doing much harm to the mean, variance or correlation information. However, a single simulation does not show much insight to a process. So, we repeat the process $S$ times and found the estimates of bias and root mean-squared error for $S=\{500,800,1000\}$ [See Table~\ref{S:Bias} and~\ref{S:RMSE}].

Note that the values for different $S$ remain more or less same which means the bias and r.m.s.e. are consistent. One can now easily observe that the moments and correlation are well-estimated in both additive noise model and this method. However the quantile estimates are improved to a remarkable extent. Even for this large value of $\sigma$, this method still gives reliable estimate for the same. However, a reliable estimate is only relevant if the data is well protected against disclosure. To measure the disclosure risk we calculate the value of the statistic $\frac{\sum_{s=1}^S{I_{[\tau_s \in (X_i-d,X_i+d)]}}}{S}$ for $\tau=Z_i$ (as shown in Table~\ref{Disclosure_Risk}), assuming $Z_i$ to be the best estimator of $X_i$. 

\begin{table}
\begin{center}
\caption{{Disclosure risk for Additive Noise Model and our method for $S=1000$ simulations for simulated data}}
{\begin{tabular}{|c|c|c| }
\hline
{d} & CM & ADM \\
 \hline
250 & 0.153 & 0.221  \\
500 & 0.298 & 0.393  \\
1000 & 0.541 & 0.632  \\
1500 & 0.712 & 0.777  \\
2000 & 0.819 & 0.864  \\
\hline
\end{tabular}} \\
{\emph{\textbf{Note}}:The simulated values were same for  $S=$ 500,800 and 1000.}
\end{center}
\label{Disclosure_Risk}
\end{table}

\noindent 
\smallskip

\noindent One can easily observe that the average number of $Z_i$ that lies within $d$-boundary of $X_i$ is very small for average $d$. Even for $d$ as large as the dispersion of $X$, it is about $55\%$ for Conditional Masking and $60\%$ for Additive Noise Model. Hence the data is well masked for both the procedures however conditional masking gives even better result than additive noise model. 

To illustrate consistency of the estimators we also simulate data-sets for increasing $n$. We calculate the bias, r.m.s.e. and disclosure risk values for $n=\{2000,5000,10000\}$ [See Table~\ref{n:Bias} and~\ref{n:RMSE}].

Looking at the tables one can easily see that the bias and r.m.s.e. for both the estimators decrease for increasing $n$ which shows numerically that the estimators are consistent.

\paragraph*{\emph{Note}} We only compare the results of Additive Noise Model (ADM) and our method but not data swapping, because a simple data swapping would result in exact estimates for mean, variance and quantiles, and complete loss of information in correlation information.

\setcounter{table}{2}
\begin{sidewaystable}[h]
\caption{{Estimated bias from original and obfuscated data using additive noise model and our method for simulated data}}
{\begin{tabular}{|c|c|c|c|c|c|c|c|c|c|c|c|c|c|}
\hline

Statistic & & 0.1 & 0.2 & 0.3 & 0.4 & 0.5 & 0.6 & 0.7 & 0.8 & 0.9 & Mean & s.d. & Cor \\
\hline
TRUE & & -1599.438 & -906.291 & -500.826 & -213.144 & 10 & 233.144 & 520.826 & 926.291 & 1619.438 & 10 & 1414.214 & -0.7 \\
\hline
\hline
\multirow{3}{*}{$T1$} 
&$S=500$  & -7.451 & 2.401 & 2.102 & -1.863 & 0.521 & 0.432 & 2.108 & 2.427 & 6.227 & 0.775 & 2.323 & 0.017 \\ \cline{2-14}
&$S=800$ & -8.471 & -2.942 & -1.879 & -3.991 & -1.4 & -0.642 & -0.442 & 0.144 & 3.795 & -1.614 & 1.474 & 0.014 \\ \cline{2-14}
&$S=1000$ & -7.352 & -1.811 & -0.905 & -3.454 & -1.258 & -0.563 & 0.247 & 1.462 & 5.063 & -0.867 & 0.823 & 0.014 \\ \cline{2-14}
\hline
\hline
\multirow{3}{*}{$Tb$} 
&$S=500$ & -45.966 & -38.447 & -38.614 & -30.765 & 0.246 & 31.645 & 41.11 & 42.056 & 44.454 & 0.775 & 2.323 & 0.017 \\ \cline{2-14}
&$S=800$ &   -47.689 & -43.033 & -41.884 & -33.153 & -1.689 & 29.832 & 39.034 & 40.259 & 42.477 & -1.614 & 1.474 & 0.014 \\ \cline{2-14}
&$S=1000$ & -46.583 & -41.949 & -41.176 & -32.686 & -1.434 & 29.986 & 39.394 & 41.137 & 44.209 & -0.867 & 0.823 & 0.014 \\ \cline{2-14}
\hline
\hline
\multirow{3}{*}{$ANM$} 
&$S=500$ & -94.454 & -80.728 & -71.085 & -45.791 & 4.934 & 54.31 & 76.414 & 87.079 & 97.813 & 3.775 & 1.188 & 0.014 \\ \cline{2-14}
&$S=800$ & -91.866 & -81.986 & -73.177 & -47.933 & 3.475 & 53.422 & 75.265 & 82.036 & 88.739 & 0.677 & 0.508 & 0.012 \\ \cline{2-14}
&$S=1000$ & -90.616 & -81.279 & -73.082 & -48.576 & 2.916 & 53.972 & 77.217 & 84.69 & 88.395 & 1.858 & 0.369 & 0.013 \\ \cline{2-14}
\hline
\end{tabular}}
\label{S:Bias}
\end{sidewaystable}

\setcounter{table}{3}
\begin{sidewaystable}[h]
\caption{{Estimated R.M.S.E. from original and obfuscated data using additive noise model and our method for simulated data}}
{\begin{tabular}{|c|c|c|c|c|c|c|c|c|c|c|c|c|c|}
\hline

Statistic & & 0.1 & 0.2 & 0.3 & 0.4 & 0.5 & 0.6 & 0.7 & 0.8 & 0.9 & Mean & s.d. & Cor \\
\hline
TRUE & & -1599.438 & -906.291 & -500.826 & -213.144 & 10 & 233.144 & 520.826 & 926.291 & 1619.438 & 10 & 1414.214 & -0.7 \\
\hline
\hline
\multirow{3}{*}{$T1$} 
&$S=500$ & 105.718 & 68.505 & 54.114 & 42.623 & 36.512 & 41.823 & 53.924 & 76.498 & 111.902 & 43.763 & 49.925 & 0.068 \\ \cline{2-14}
&$S=800$ & 107.288 & 71.566 & 54.496 & 43.241 & 36.921 & 43.596 & 54.859 & 75.663 & 112.72 & 45.169 & 50.679 & 0.068 \\ \cline{2-14}
&$S=1000$ & 107.782 & 72.018 & 55.38 & 43.688 & 37.324 & 43.612 & 54.631 & 75.574 & 111.266 & 45.644 & 51.006 & 0.068 \\ \cline{2-14}
\hline
\hline
\multirow{3}{*}{$Tb$} 
&$S=500$ & 104.803 & 76.242 & 63.377 & 51.001 & 36.652 & 49.97 & 62.705 & 77.296 & 107.898 & 45.169 & 50.679 & 0.068 \\ \cline{2-14}
&$S=800$ & 104.803 & 76.242 & 63.377 & 51.001 & 36.652 & 49.97 & 62.705 & 77.296 & 107.898 & 45.169 & 50.679 & 0.068 \\ \cline{2-14}
&$S=1000$ & 105.643 & 76.396 & 63.453 & 51.097 & 36.886 & 50.12 & 62.905 & 77.537 & 107.897 & 45.644 & 51.006 & 0.068 \\ \cline{2-14}
\hline
\hline
\multirow{3}{*}{$ANM$} 
&$S=500$ & 190.174 & 134.445 & 105.501 & 78.966 & 62.894 & 85.24 & 108.607 & 138.752 & 190.351 & 43.877 & 57.674 & 0 \\ \cline{2-14}
&$S=800$ & 191.034 & 134.439 & 107.09 & 80.88 & 62.741 & 83.996 & 106.855 & 136.38 & 185.941 & 45.013 & 57.753 & 0 \\ \cline{2-14}
&$S=1000$ & 192.051 & 133.318 & 106.236 & 81.216 & 62.656 & 85.37 & 109.275 & 136.992 & 186.095 & 44.98 & 57.573 & 0 \\ \cline{2-14}
\hline
\end{tabular}}
\label{S:RMSE}
\end{sidewaystable}
 
\setcounter{table}{4}
\begin{sidewaystable}[h]
\caption{{Estimated bias from obfuscated data using additive noise model and our method for simulated data}}
{\begin{tabular}{|c|c|c|c|c|c|c|c|c|c|c|c|c|c|}
\hline

Statistic & & 0.1 & 0.2 & 0.3 & 0.4 & 0.5 & 0.6 & 0.7 & 0.8 & 0.9 & Mean & s.d. & Cor \\
\hline
Original & & -1599.438 & -906.291 & -500.826 & -213.144 & 10 & 233.144 & 520.826 & 926.291 & 1619.438 & 10 & 1414.14 & -0.7 \\
\hline
\hline
\multirow{3}{*}{$T1$} 
&$n=2000$ & -7.352 & -1.811 & -0.905 & -3.454 & -1.258 & -0.563 & 0.247 & 1.462 & 5.063 & 18.791 & -21.263 & 0.04 \\ \cline{2-14}
&$n=5000$ & -1.588 & 0.801 & 0.751 & 0.796 & 1.349 & 0.745 & 1.43 & 0.945 & 1.799 & -15.926 & 8.02 & 0.068 \\ \cline{2-14}
&$n=10000$ & 0.795 & 1.277 & 0.978 & 0.582 & 0.542 & 0.633 & 0.301 & 1.147 & 1 & 19.138 & 15.358 & 0.045\\ \cline{2-14}
\hline
\hline
\multirow{3}{*}{$Tb$} 
&$n=2000$ & -46.583 & -41.949 & -41.176 & -32.686 & -1.434 & 29.986 & 39.394 & 41.137 & 44.209 & 18.791 & -21.263 & 0.04 \\ \cline{2-14}
&$n=5000$ & -28.755 & -27.651 & -27.321 & -22.492 & 1.039 & 24.445 & 29.197 & 29.306 & 29.454 & -15.926 & 8.02 & 0.068 \\ \cline{2-14}
&$n=10000$ & -20.704 & -20.347 & -20.296 & -17.855 & 0.67 & 19.148 & 21.835 & 22.358 & 22.539 & 19.138 & 15.358 & 0.045 \\ \cline{2-14}
\hline
\hline
\multirow{3}{*}{$ANM$} 
&$n=2000$ & 0.795 & 1.277 & 0.978 & 0.582 & 0.542 & 0.633 & 0.301 & 1.147 & 1 & 19.138 & 15.358 & 0.045 \\ \cline{2-14}
&$n=5000$ & -20.704 & -20.347 & -20.296 & -17.855 & 0.67 & 19.148 & 21.835 & 22.358 & 22.539 & 19.138 & 15.358 & 0.045 \\ \cline{2-14}
&$n=10000$ & -41.152 & -42.353 & -38.741 & -30.838 & 0.355 & 31.815 & 40.871 & 44.796 & 40.295 & 18.227 & 16.06 & 0 \\ \cline{2-14}
\hline
\end{tabular}}
\label{n:Bias}
\end{sidewaystable}

\setcounter{table}{5}
\begin{sidewaystable}[h]
\caption{{Estimated R.M.S.E. from obfuscated data using additive noise model and our method for simulated data}}
{\begin{tabular}{|c|c|c|c|c|c|c|c|c|c|c|c|c|c|}
\hline

Statistic & & 0.1 & 0.2 & 0.3 & 0.4 & 0.5 & 0.6 & 0.7 & 0.8 & 0.9 & Mean & s.d. & Cor \\
\hline
Original & & -1599.438 & -906.291 & -500.826 & -213.144 & 10 & 233.144 & 520.826 & 926.291 & 1619.438 & 10 & 1414.14 & -0.7 \\
\hline
\hline
\multirow{3}{*}{$T1$} 
&$n=2000$ & 107.782 & 72.018 & 55.38 & 43.688 & 37.324 & 43.612 & 54.631 & 75.574 & 111.266 & 45.644 & 418.158 & 0.068 \\ \cline{2-14}
&$n=5000$ & 68.685 & 46 & 35.846 & 27.968 & 23.005 & 27.244 & 34.355 & 45.338 & 68.205 & 28.834 & 415 & 0.045 \\ \cline{2-14}
&$n=10000$ &47.506 & 32.43 & 24.543 & 20.041 & 16.651 & 19.614 & 25.245 & 33.946 & 49.787 & 20.032 & 414.052 & 0.033\\ \cline{2-14}
\hline
\hline
\multirow{3}{*}{$Tb$} 
&$n=2000$ & 105.643 & 76.396 & 63.453 & 51.097 & 36.886 & 50.12 & 62.905 & 77.537 & 107.897 & 45.644 & 418.158 & 0.068 \\ \cline{2-14}
&$n=5000$ & 66.908 & 49.649 & 41.818 & 33.894 & 23.114 & 34.968 & 42.638 & 50.547 & 68.84 & 28.834 & 415 & 0.045 \\ \cline{2-14}
&$n=10000$ & 47.627 & 35.585 & 30.165 & 25.429 & 16.634 & 26.322 & 31.575 & 38.226 & 50.299 & 20.032 & 414.052 & 0.033\\ \cline{2-14}
\hline
\hline
\multirow{3}{*}{$ANM$} 
&$n=2000$ & 192.051 & 133.318 & 106.236 & 81.216 & 62.656 & 85.37 & 109.275 & 136.992 & 186.095 & 44.98 & 418.561 & 0\\ \cline{2-14}
&$n=5000$ & 137.767 & 92.931 & 75.809 & 60.332 & 42.601 & 59.007 & 77.473 & 94.47 & 141.541 & 28.109 & 415.072 & 0 \\ \cline{2-14}
&$n=10000$ & 112.455 & 75.626 & 60.848 & 49.98 & 35.015 & 49.508 & 62.339 & 78.12 & 111.537 & 19.876 & 414.997 & 0 \\ \cline{2-14}
\hline
\end{tabular}}
\label{n:RMSE}
\end{sidewaystable}

\section{Real-Life Data}
We also consider a real-life example scenario to check the application of the discussed procedure. We collect a data-set of 1st and 2nd semester marks achieved by $445$ students in the M.Stat 2nd yr program of Indian Statistical Institute Kolkata over 10years 2006-2015. Now since marks is a sensitive data, it cannot be released in its raw form. So we apply the above problem to this data and try to find the results. Standard variation of the data was checked to be approximately $100$; so for obfuscating distribution we chose $\sigma=100$. 

\begin{figure}[!h]
\centering
\includegraphics[scale=0.3]{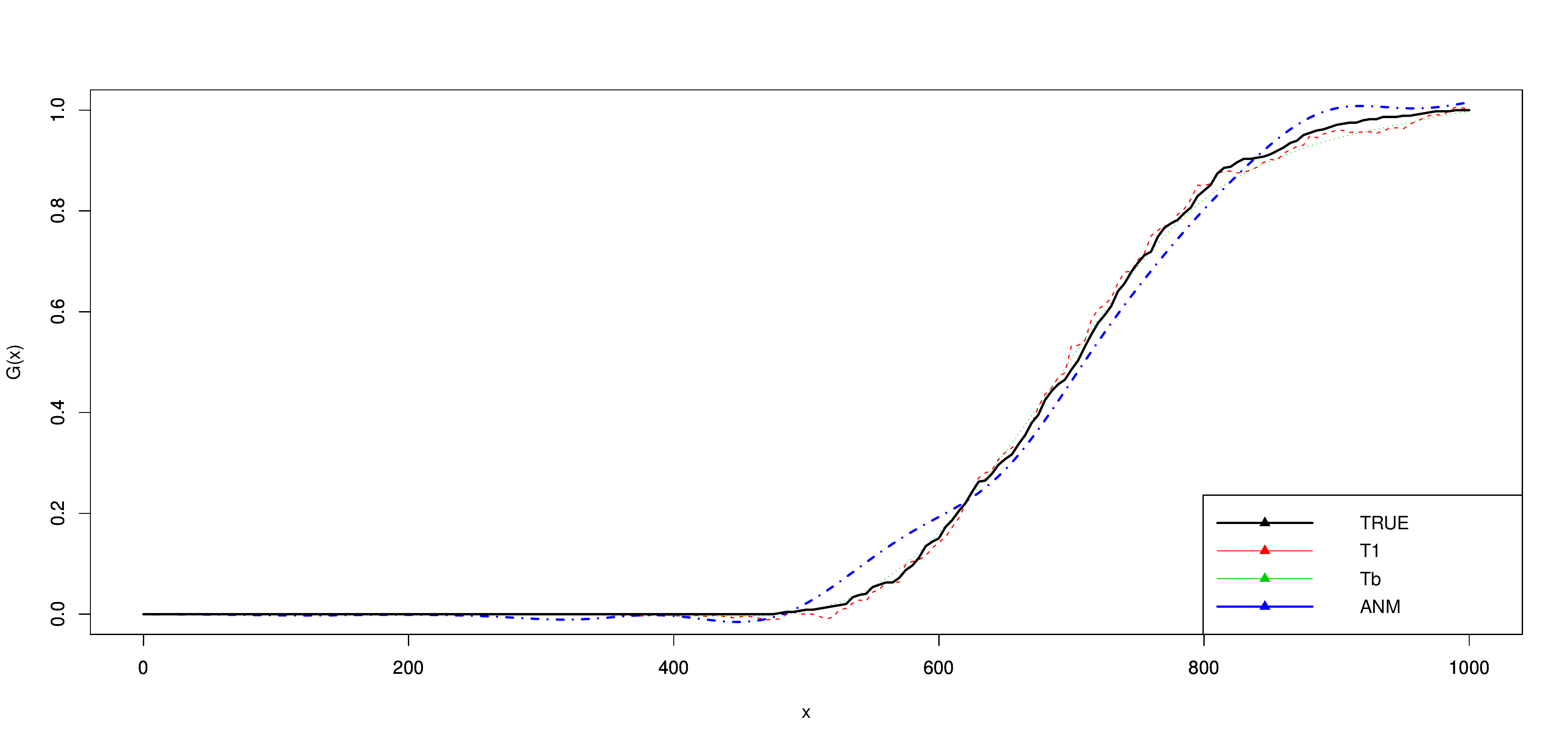}
\caption{{Estimated distribution curve for the data set of 445 students from true data set and obfuscated data-sets using Additive Noise Model and our method}}
\label{graph:real}
\end{figure}

A problem we found while applying the procedure was that since the data points are integers, the obfuscated points that are swapped are integers while those with added noise had decimal parts. So, one can clearly see which values are swapped and which are added noise to. To avoid this, we only considered the nearest integer to the Normal noise added instead of the exact noise and carried the same procedure. The results are given below. Table \ref{Masking_check} represents true and obfuscated values of 10 data points to show how the values are masked. Then from the obfuscated values the true distribution and quantiles are estimated as shown in Fig \ref{graph:real} and Table \ref{Estimates:real} respectively. Since $n$ is moderate here, we chose $p=0.55$.

\begin{table}
\begin{center}
\caption{{ True and Obfuscated Values of 10 data points selected from the list of 445 students}}
{\begin{tabular}{|c|c|c|c| }
\hline
Point & True & CM & ADM \\
\hline
"1" & 761 & 671 & 760.422 \\
"2" & 856 & 669 & 770.865 \\
"3" & 808 & 748 & 816.227 \\
"4" & 880 & 720 & 885.171 \\
"5" & 933 & 781 & 807.114 \\
"6" & 946 & 498 & 988.723 \\
"7" & 901 & 737 & 819.712 \\
"8" & 791 & 767 & 855.312 \\
"9" & 739 & 937 & 775.343 \\
"10" & 720 & 625 & 638.888 \\
\hline
\end{tabular} 
}
\label{Masking_check}
\end{center}
\end{table}

\begin{table}
\begin{center}
\caption{ Estimates of different Statistics from original and obfuscated data-sets corresponding to 445 students}
{
\begin{tabular}{|c|c|c|c|c| }
\hline
Statistic & Original & T1 & Tb & ANM \\
\hline
"0.1" & 580.8 & 577 & 574.716 & 543.663 \\
"0.2" & 612.8 & 617 & 613.072 & 604.883 \\
"0.3" & 645.2 & 645 & 643.845 & 654.71 \\
"0.4" & 675.6 & 673 & 671.655 & 684.003 \\
"0.5" & 700 & 699 & 698.033 & 709.56 \\
"0.6" & 727 & 720 & 724.572 & 736.245 \\
"0.7" & 750 & 750 & 753.56 & 765.846 \\
"0.8" & 786 & 782 & 789.721 & 798.864 \\
"0.9" & 826.6 & 850 & 850.935 & 836.601 \\
"Mean" & 703.047 & 706.189 & 706.189 & 701.878 \\
"s.d." & 97.951 & 97.048 & 97.048 & 71.094 \\
"Cor" & 0.68 & 0.603 & 0.603 & 0.828 \\
\hline
\end{tabular} 
}
\label{Estimates:real}
\end{center}
\end{table}

We can see that the process works well for real-data also.
\section{Conclusion}
This way of masking numerical data with a random method can be useful in various applications because it retains much of quantile and moment information even after obfuscating a data-set sufficiently, with not a huge loss in its correlation information with other non-sensitive variables. However, even if there are more than one sensitive variables, the Bernoulli variable should be chosen only once; if not, the correlation between the two sensitive variables may be hard to get back from the obfuscated data-set. If multiple sensitive attributes are present in a data-set, and some or few of them are obfuscated with this method, then studying correlation between the obfuscated variables can be an interesting problem for future work.

In our previous work, we had used \emph{Laplace Error} to obfuscate the variables, but here we use \emph{Normal Error} because \emph{lemma \ref{Lem:Norm}} would not hold for Laplace Error and it would become hard to get back the quantiles. Infact, any other Error distribution other than Normal may make it very hard to get back the quantiles.

Even after getting unbiased estimation for the distribution curve, the problem of performing a statistical hypothesis testing or getting a confidence interval is still a problem of concern because, even here, the variance is not easily estimable. 

However, the estimation process works good for this method and gives very good estimates for large sample sizes.

\section{Appendix}
\subsection{\underline{Proof of Theorem 2.1}}
\begin{proof}

To find the raw moments of $X$, denoted by $\mu_{(X,k)}$, in terms of moments of $Z$, we first need to check that the moments of $Z$ exists whenever that of $X$ does. $k^{th}$ absolute raw moment of $Z_i$ is given by
\medskip

\noindent
$\begin{array}{ll}
E[|Z_i^k|] &= E[|Z_i|^k \given B_i=1]\cdot P[B_i=1] + E[|Z_i|^k \given B_i=0]\cdot P[B_i=0]\\
&= p\cdot \frac{1}{n-1}\sum_{j=1,j \neq i}^{n}E[|X_{j}|^k \given j \mbox{ is chosen}] + (1-p)\cdot E[|X_i+Y_i|^k] \\
&= p\cdot  \frac{1}{n-1}\sum_{j=1,j \neq i}^{n}{E[|X_j|^k]} + (1-p)\cdot E[|X_i+Y_i|^k] \\
&\leq p\cdot {E[|X_1|^k]} + (1-p)\cdot E[|X_1|^k+|Y_1|^k] , \\
& \mbox{ [by Minkowski's Inequality and the fact that $X_i$ and $Y_i$'s are i.i.d.]} \\
&< \infty
\end{array}$

since, $E|X_j|^k <\infty$, and $E|Y_j|^k =\frac{\sigma^k 2^\frac{k}{2}\Gamma(\frac{k+1}{2})}{\sqrt{\pi}} < \infty$ $\forall k \in \mathbb{N}$.

Thus, the moments of $Z$ exists if that of $X$ exists. Now, we try to find the estimates of moments of $X$ in terms of moments of $Z$.
\medskip

\noindent
$\begin{array}{ll}
\mu_{(Z,k)}  &= E[Z_i^k] \\
&= E[Z_i^k \given B_i=1]\cdot P[B_i=1] + E[Z_i^k \given B_i=0]\cdot P[B_i=0]\\
&= p\cdot \frac{1}{n-1}\sum\limits_{j=1,j \neq i}^{n}{E [X_j^k]} + (1-p)\cdot E[(X_i+Y_i)^k] \\
&= E[X_i^k]+ (1-p)\{ k.E[X_i^{k-1}]\cdot E[Y_i] + \ldots + E[Y_i^k]\} \\
&= \mu_{(X,k)} + (1-p)\cdot \sum\limits_{\substack{j=2 \\ j \mbox{ \scriptsize is even }}}^{2[\frac{k}{2}]}{{k\choose j}\mu_{(X,k-j)}\mu_{(Y,j)}} 
\end{array}$
\smallskip

The above equation follows since odd order moment of $Y_i$ is zero. 
\smallskip

\noindent $\therefore \mu_{(X,k)}=\mu_{(Z,k)} -(1-p)\cdot \sum\limits_{j=1}^{[\frac{k}{2}]}{{k\choose 2j} {\mu_{(X,k-2j)}\mu_{(Y,2j)}}}$.

Now, $E[Z]=p\cdot E[X] + (1-p) \cdot E[X+Y] =E[X]$.
$\therefore E[\frac{1}{n}\sum_{j=1}^n{Z_j}]=E[Z]=E[X]$.

$\therefore \frac{1}{n}\sum_{j=1}^n{Z_j}$ is an unbiased estimator of $E[X]$.

Define, in general, $$ \hat{\mu}_{(X,k)}= \frac{1}{n}\sum_{j=1}^n{Z_j^k} - (1-p)\cdot \sum\limits_{j=1}^{[\frac{k}{2}]}{{k\choose 2j} \mu_{(X,k-2j)}\mu_{(Y,2j)}} \mbox{ , $\mu_{(Y,2j)}=\sigma^{2j}\frac{2^j\Gamma(j+\frac{1}{2})}{\sqrt{\pi}}<\infty$}$$

If $E[\hat{\mu}_{(X,j)}]=\mu_{(X,j)}$ , $\forall j=1,2, \ldots k$ then,

$$ E[\hat{\mu}_{(X,k+1)}]={\mu}_{(Z,k+1)} - (1-p)\cdot \sum_{j=1}^{[\frac{k}{2}]}{\mu_{(X,k-2j)}\mu_{(Y,2j)}}={\mu}_{(X,k+1)}$$

Thus, by induction, $\hat{\mu}_{(X,k)}$ is an unbiased estimator for ${\mu}_{(X,k)}$.

Also, note that $\hat{\mu}_{(X,k)}=\frac{1}{n}\sum_{j=1}^n{f(Z_j)}$ , where $f(.)$ is a polynomial function of finite degree. Also $Var(Z_j)=E(Z^{2j})-E(Z_j)^2 < \infty$ if the moments of $X$ and hence $Z$ exist and is finite $\forall n \in \mathbb{N}$. Thus $Var(\hat{\mu}_{(X,k)})=O(\frac{1}{n})$ and $\hat{\mu}_{(X,k)}$ is also a consistent estimator of $\mu_{(X,k)}$. 
\end{proof}

\noindent Now, we state and prove the following lemma, which we will require while proving Theorem 2.2 and 2.3 in subsequent sections. 

\begin{lem}
For $(x_1,x_2,x_3) \in \mathbb{R}^3$ and $(\sigma_1,\sigma_2) \in \mathbb{R^+}^2$,
$$\int_{-\infty}^{\infty}{ \phi_{\sigma_1}{(x_1-x_2)}\cdot \phi_{\sigma_2}{(x_2-x_3)} dx_2}=\phi_{\sqrt{\sigma_1^2+ \sigma_2^2}}{(x_1-x_3)}$$
\label{Lem:Norm}
\end{lem}
where $\phi_\sigma(x)=$ normal density at $x \in \mathbb{R}$ for mean $0$ and standard deviation $\sigma$.
\begin{proof}
To prove the given lemma we consider, \\

$\begin{array}{ll}
L.H.S. &=\frac{1}{2\pi\sigma_1\sigma_2}\cdot \int_{-\infty}^{\infty}{e^{{-}\frac{1}{2}\cdot[(\frac{x_1-x_2}{\sigma_1})^2+(\frac{x_2-x_2}{\sigma_2})^2]} dx_2} \\
&=\frac{1}{2\pi\sigma_1\sigma_2}\cdot \int_{-\infty}^{\infty}{e^{{-}\frac{1}{2}\cdot[x_2^2(\frac{1}{\sigma_1^2}+\frac{1}{\sigma_2^2})-2x_2 (\frac{x_1}{\sigma_1^2}+\frac{x_3}{\sigma_2^2})+\frac{x_1^2}{\sigma_1^2}+\frac{x_3^2}{\sigma_2^2}]}dx_2} \\
&=\frac{1}{\sqrt{2\pi}\sqrt{\sigma_1^2+\sigma_2^2}}\cdot e^{{-}\frac{1}{2(\frac{1}{\sigma_1^2}+\frac{1}{\sigma_2^2})}[(\frac{x_1}{\sigma_1^2}+\frac{x_3}{\sigma_2^2})^2 -(\frac{x_1}{\sigma_1^2}+\frac{x_3}{\sigma_2^2})\cdot(\frac{1}{\sigma_1^2}+\frac{1}{\sigma_2^2})]} \\
& \hspace*{25mm} \cdot \int_{-\infty}^{\infty}{\phi_{\frac{1}{\frac{1}{\sigma_1^2}+\frac{1}{\sigma_2^2}}}(x_2-(\frac{\frac{x_1}{\sigma_1^2}+\frac{x_3}{\sigma_2^2}}{\frac{1}{\sigma_1^2}+\frac{1}{\sigma_2^2}}))dx_2} \\
&=\frac{1}{\sqrt{2\pi}\sqrt{\sigma_1^2+\sigma_2^2}}\cdot e^{-\frac{(x_1-x_3)^2}{2(\sigma_1^2+\sigma_2^2)}} \\
&=R.H.S.
\end{array}$ \\

\end{proof}

\subsection{\underline{Proof of Theorem 2.2}}
\begin{proof}
Let $H(.)$ be the c.d.f. of $Z$. Then, \\
$\begin{array}{ll}
H(x) &= P[Z_i \leq x] \\
&=P[Z_i \leq x \given B_i=1]\cdot P[B_i=1] + P[Z_i \leq x \given B_i=0]\cdot P[B_i=0] \\
&=p \cdot G(x) + (1-p)\cdot \int_{-\infty}^{\infty}{\phi_\sigma(x-y)G(y)dy} \\
\end{array}$ \\
If $\tilde{G}(x)=p\cdot G(x)$, then we have the equation, \\
\begin{equation} 
\tilde{G}(x)= H(x) + \lambda \int_{-\infty}^{\infty}{\phi_\sigma(x-y)\tilde{G}(y)dy}
\label{Eqn:Fred}
\end{equation}
 To find a solution to Equation~\eqref{Eqn:Fred}, note that this is a Fredholm Equation of second kind and a solution for such a problem for $|\lambda|<1$ is given by the famous Liouville-Neumann Series,
\begin{equation}
\tilde{G}(x)=\underset{n \rightarrow \infty}{lim}\sum_{t=0}^n{\lambda^t u_t(x)} \mbox{ , where, $u_o(x)=H(x)$ ,}
\label{Soln:Fred}
\end{equation} 
 
 $u_t(x)=\int_{-\infty}^{\infty} \ldots \int_{-\infty}^{\infty} {\phi_\sigma(x-x_1)\cdot \phi_\sigma(x_2-x_1)\cdots \phi_\sigma(x_{t-1}-x_t)H(x_t)dx_1 dx_2 \ldots dx_t}$
 
 Now $H(x)$ is unknown, so we use $\hat{H}(x)=\frac{1}{n}\sum_{j=1}^n{\mathbb{I}_{[Z_j \leq x]}}$ instead. Here, $\mathbb{I}_A$ is the indicator function for event $A$. \\
 \smallskip
 For $t \in \mathbb{N}$, the integrand in $\hat{u}_t(x)$ is integrable since, \\
 $\begin{array}{ll}
 |\hat{u}_t(x)| &= |\int_{-\infty}^{\infty} \ldots \int_{-\infty}^{\infty} {\phi_\sigma(x-x_1)\cdot \phi_\sigma(x_2-x_1)\cdots \phi_\sigma(x_{t-1}-x_t)\hat{H}(x_t)dx_1 dx_2 \ldots dx_t}| \\
 &\leq \int_{-\infty}^{\infty} \ldots \int_{-\infty}^{\infty} |{\phi_\sigma(x-x_1)\cdot \phi_\sigma(x_2-x_1)\cdots \phi_\sigma(x_{t-1}-x_t)\hat{H}(x_t)|dx_1 dx_2 \ldots dx_t} \\
 &\leq \int_{-\infty}^{\infty} \ldots \int_{-\infty}^{\infty} {\phi_\sigma(x-x_1)\cdot \phi_\sigma(x_2-x_1)\cdots \phi_\sigma(x_{t-1}-x_t)dx_1 dx_2 \ldots dx_t} \\
 &= \int_{-\infty}^{\infty}{\phi_{\sigma\sqrt{t}}(x-x_t)dx_t} \mbox{ , by lemma 5.1} \\
 &=1 < \infty \mbox{ , $\forall x \in \mathbb{R}$}
 \end{array}$ \\
 Also, \\
 
 $\begin{array}{ll}
 \hat{u}_t(x) &= \int_{-\infty}^{\infty} \ldots \int_{-\infty}^{\infty} {\phi_\sigma(x-x_1)\cdot \phi_\sigma(x_2-x_1)\cdots \phi_\sigma(x_{t-1}-x_t)\hat{H}(x_t)dx_1 dx_2 \ldots dx_t} \\
 &=\int_{-\infty}^{\infty}{\phi_{\sigma\sqrt{t}}(x-x_t)\hat{H}(x_t)dx_t} \mbox{ , by lemma 5.1} \\
 &=\frac{1}{n}\sum_{j=1}^n{\int_{Z_j}^{\infty}{\phi_{\sigma\sqrt{t}}(x-x_t)dx_t}} \\
 &=\frac{1}{n}\sum_{j=1}^n{\int_{-\infty}^{x-Z_j}{\phi_{\sigma\sqrt{t}}(y)dy}} \mbox{ , Taking $y=x-x_t$} \\ 
 &=\frac{1}{n}\sum_{j=1}^n{\Phi_{\sigma\sqrt{t}}(x-Z_j)}
  \end{array}$ \\
  
  Thus $\hat{\tilde{G}}(x)= \sum_{t=0}^\infty{\lambda^t \hat{u}_t(x)}= \frac{1}{n}\sum_{j=1}^n {\sum_{t=0}^\infty{\lambda^t \Phi_{\sigma\sqrt{t}}(x-Z_j)}}$ is an estimator of $\tilde{G}(x)$. \\
  
  $\begin{array}{ll}
 E[\hat{\tilde{G}}(x)] &=\sum_{t=0}^\infty{\lambda^t E[\hat{u}_t(x)]} \\
 &= \sum_{t=0}^\infty{\lambda^t E[\int_{-\infty}^{\infty} \ldots \int_{-\infty}^{\infty} {\phi_\sigma(x-x_1)\cdots \phi_\sigma(x_{t-1}-x_t)\hat{H}(x_t)dx_1 \ldots dx_t}]} \\
 &= \sum_{t=0}^\infty{\lambda^t E[\int_{-\infty}^{\infty}{\phi_{\sigma\sqrt{t}}(x-x_t)\hat{H}(x_t)dx_t}]} \\
 &=  \sum_{t=0}^\infty{\lambda^t \int_{-\infty}^{\infty}{\phi_{\sigma\sqrt{t}}(x-x_t)E[\hat{H}(x_t)]dx_t}} \mbox{ , by Tonelli's Theorem} \\
 &=  \sum_{t=0}^\infty{\lambda^t \int_{-\infty}^{\infty}{\phi_{\sigma\sqrt{t}}(x-x_t)H(x_t)dx_t}} \\
 &= \sum_{t=0}^\infty{\lambda^t \int_{-\infty}^{\infty} \ldots \int_{-\infty}^{\infty} {\phi_\sigma(x-x_1)\cdots \phi_\sigma(x_{t-1}-x_t)H(x_t)dx_1 \ldots dx_t}} \\
 &= \tilde{G}(x)
  \end{array}$ 
  
  Thus $\frac{\hat{G}(x)}{p}$ is an unbiased estimator for $G(x)$

\end{proof}

\subsection{\underline{Proof of Theorem 2.3}}
\begin{proof}
Note that in Equation~\eqref{Eqn:Fred}, the integral term is a convolution of two functions and hence can be easily interchanged. Thus, taking derivatives on both sides we have,
\begin{equation} 
\tilde{g}(x)= h(x) + \lambda \int_{-\infty}^{\infty}{\phi_\sigma(x-y)\tilde{g}(y)dy} 
\label{Eqn:pdf}
\end{equation}
where $\tilde{g}(x)=\frac{d}{dx}\{\tilde{G}(x)\}=pg(x)$ and $g(x)$, $h(x)$ are density of $X$ and $Z$ respectively.

To find a solution to Equation~\eqref{Eqn:pdf}, note that this is a Fredholm Equation of second kind and a solution for such a problem for $|\lambda|<1$ is given by the Liouville-Neumann Series,
 $$ \tilde{g}(x)=\underset{n \rightarrow \infty}{lim}\sum_{t=0}^n{\lambda^t u_t(x)} \mbox{ , where, $u_o(x)=h(x)$ ,}$$
 
 $u_t(x)=\int_{-\infty}^{\infty} \ldots \int_{-\infty}^{\infty} {\phi_\sigma(x-x_1)\cdot \phi_\sigma(x_2-x_1)\cdots \phi_\sigma(x_{t-1}-x_t)h(x_t)dx_1 dx_2 \ldots dx_t}$
 
 Now $h(x)$ is unknown, so we use $\hat{h}(x)=\frac{1}{n}\sum_{j=1}^{nb}{K(\frac{x-Z_j}{b})}$ instead. Here, $K(.)$ is the Gaussian Kernel function and $b$ is the bandwidth selected by Silvermans rule of thumb. \\
 For $t \in \mathbb{N}$, the integrand in $\hat{u}_t(x)$ is integrable in a similar reason as in the proof of Theorem 2.1. Also,

$\begin{array}{ll}
 \hat{u}_t(x) &= \int_{-\infty}^{\infty} \ldots \int_{-\infty}^{\infty} {\phi_\sigma(x-x_1)\cdot \phi_\sigma(x_2-x_1)\cdots \phi_\sigma(x_{t-1}-x_t)\hat{h}(x_t)dx_1 dx_2 \ldots dx_t} \\
 &=\int_{-\infty}^{\infty}{\phi_{\sigma\sqrt{t}}(x-x_t)\hat{h}(x_t)dx_t} \mbox{ , by lemma 5.1} \\
 &=\frac{1}{n}\sum_{j=1}^n{\int_{-\infty}^{\infty}{\phi_{\sigma\sqrt{t}}(x-x_t)\phi_{b}(x_t-Z_j)dx_t}} \\
 &=\frac{1}{n}\sum_{j=1}^n{\phi_{\sqrt{t\sigma^2 + b^2}}(x-Z_j)} \mbox{ , by lemma 5.1} \\ 
 
  \end{array}$ \\
  
  Thus $\hat{\tilde{g}}(x)= \sum_{t=0}^\infty{\lambda^t \hat{u}_t(x)}= \frac{1}{n}\sum_{j=1}^n {\sum_{t=0}^\infty{\lambda^t \phi_{\sqrt{t\sigma^2 + b^2}}(x-Z_j)}}$ is an estimator of $\tilde{g}(x)$ and hence $$ \hat{G}(x)= \frac{1}{np}\sum_{j=1}^{n} {\sum_{t=0}^\infty{\lambda^t \Phi_{\sqrt{t\sigma^2 + b^2}}(x-Z_j)}} $$ is an estimator for $G(x)$.\\
  
  Note that this estimator is a linear series of normal c.d.f.s each with positive variance and hence are smooth functions, which makes the function $\hat{G}(x)$, a smooth function. \\

\end{proof}
 
\subsection{\underline{Proof of Theorem 2.4}}
\begin{proof}

We have \ $T_i(x)=\frac{1}{np}\sum_{j=1}^n {\sum_{t=0}^\infty{\lambda^t \Phi_{\sigma_t}(x-Z_j)}}$ where $\sigma_t=\sigma\sqrt{t}$ for $T_1(x)$ and $\sqrt{t\sigma^2 + b^2}$ for $T_b(x)$. Since, $\Phi_{\sigma_t}(x-Z_j) \leq 1$ $\forall t \in \mathbb{N}$ and $\sum_{t=0}^\infty{\lambda^t}=p$, $\frac{1}{p}\sum_{t=0}^\infty{\lambda^t \Phi_{\sigma_t}(x-Z_j)} = T_{ij} \mbox{ ( say)}\leq 1$.

 $T_i(x)=\frac{1}{n}\sum_{j=1}^n{T_{ij}}$, i.e., average of $n$ i.i.d. random variables each of whose value is less than or equal to $1$. Thus,
$$Var(T_i(x)) = \frac{1}{n}\cdot Var(T_{i1}(x)) \leq \frac{1}{n}E(T_{i1}^2) = O(\frac{1}{n})$$
Since $T_1(x)$ is unbiased, the last equation implies convergence in probability of $T_1(x)$ to its expected value $G(x)$. For the smooth estimator,

$$E(T_b(x))=E[\frac{1}{np}\sum_{j=1}^n{\sum_{t=0}^\infty{\lambda^t\cdot \Phi_{b_t}(x-Z_j)}}]=\frac{1}{p}\sum_{t=0}^\infty{\lambda^t\int_{-\infty}^{\infty}{\Phi(\frac{x-z}{\sqrt{t\sigma^2 +b^2}})h(z)dz}}$$

The true c.d.f. can be written as, (using Equation~\ref{Eqn:Fred})

$\begin{array}{ll}
G(x)&= \frac{1}{p} \sum\limits_{t=0}^\infty{\lambda^t \int_{-\infty}^{\infty} \ldots \int_{-\infty}^{\infty} {\phi_\sigma(x-x_1)\cdots \phi_\sigma(x_{t-1}-x_t)H(x_t)dx_1 \ldots dx_t}} \\
&= \frac{1}{p} \sum\limits_{t=0}^\infty{\lambda^t \int_{-\infty}^{\infty}  {\phi_{\sqrt{t\sigma^2}}(x-x_t)\cdot H(x_t)dx_t}} \\
&= \frac{1}{p}\sum\limits_{t=0}^\infty{\lambda^t\int_{-\infty}^{\infty}{\Phi_{\sqrt{t\sigma^2}}(x-x_t)h(x_t)dx_t}} \\
&= \frac{1}{p}\sum\limits_{t=0}^\infty{\lambda^t\int_{-\infty}^{\infty}{\Phi(\frac{x-x_t}{\sqrt{t\sigma^2}})h(x_t)dx_t}} 
\end{array}$ 
\smallskip

\noindent Since $\int_{-\infty}^{\infty}{\phi_{\sqrt{t\sigma^2}}(x-x_t)\cdot H(x_t)dx_t}=\int_{-\infty}^{\infty}{\Phi_{\sqrt{t\sigma^2}}(x-x_t)\cdot h(x_t)dx_t}$ both being c.d.f. of $Z+N(0,\sqrt{t\sigma^2})$.

Thus, we have an expression for the bias at point $x$, denoted by $B(x)$, as given below.
\smallskip

\noindent $\begin{array}{ll}
|B(x)| &= |\frac{1}{p}\sum\limits_{t=0}^\infty{\lambda^t\int_{-\infty}^{\infty}[{{\Phi(\frac{x-z}{\sqrt{t\sigma^2 +b^2}})}-\Phi(\frac{x-z}{\sqrt{t\sigma^2}})}h(z)]dz}| \\
&\leq \frac{1}{p}| \underbrace{\int_{-\infty}^{\infty}{\mathbb{I}(x-z)h(z)dz}-\int_{-\infty}^{\infty}{\Phi(\frac{x-z}{b})h(z)dz}|}_{\textsf{Tr}_1} \\
& ~+ \frac{1}{p} \underbrace{\sum_{t=1}^\infty{|\lambda|^t\int_{-\infty}^{\infty}{|{\Phi(\frac{x-z}{\sqrt{t\sigma^2 +b^2}})}-\Phi(\frac{x-z}{\sqrt{t\sigma^2}})|}h(z)dz}}_{\textsf{Tr}_2} 
\end{array}$

$\begin{array}{ll}
\textsf{Tr}_1 &= |\int_{-\infty}^{\infty}{\mathbb{I}(x-z)h(z)dz}-\int_{-\infty}^{\infty}{\Phi(\frac{x-z}{b})h(z)dz}| \\
&= |H(x)- \int_{-\infty}^{\infty}{\Phi(x-z,0,b)h(z)dz}| \\
&= |H(x)-H^{\star}(x)|
\end{array}$

where $H^{\star}(x)$ is the c.d.f. of $Z+N(0,b^2)$, $Z$ is a r.v. with p.d.f. $h(.)$. As $n \rightarrow \infty$, $b \rightarrow 0$ and $N(0,b^2) \overset{P}{\rightarrow} 0$. Thus, by Slutsky's theorem, $Z+N(0,b^2) \Longrightarrow Z$ in distribution as $b \rightarrow 0$, or, $|H(x)-H^{\star}(x)| \rightarrow 0$ as $n \rightarrow \infty$.
\medskip

\noindent For $\textsf{Tr}_2$, since $\Phi(\frac{x-z}{\sqrt{y}})$ is a continuous differentiable function in $y$, expanding the function by Taylor Series at $y=t\sigma^2$ we have $|{\Phi(\frac{x-z}{\sqrt{t\sigma^2 +b^2}})}-\Phi(\frac{x-z}{\sqrt{t\sigma^2}})|=|(t\sigma^2+b^2 - t\sigma^2)(\phi(\frac{x-z}{\sqrt{y*}})\frac{x-z}{y*^{3/2}})|$ where $y^*$ is a point between $t\sigma^2$ and $t\sigma^2 +b^2$. Also, it is easy to show that $|x\phi(x)| \leq 1$ for all $x \in \mathcal{R}$( as discussed in {\em Note 5.1}), which implies $|\phi(\frac{x-z}{\sqrt{y*}})\frac{x-z}{y*^{1/2}}| \leq 1$.  Thus,
\medskip

\noindent $\begin{array}{ll}
\textsf{Tr}_2 &= \sum_{t=1}^\infty{|\lambda|^t\int_{-\infty}^{\infty}{|{\Phi(\frac{x-z}{\sqrt{t\sigma^2 +b^2}})}-\Phi(\frac{x-z}{\sqrt{t\sigma^2}})|}h(z)dz} \\
&\leq \sum_{t=1}^\infty{|\lambda|^t \frac{b^2}{t \sigma^2}} \\
&\leq \frac{b^2}{\sigma^2} \sum_{t=1}^\infty{\frac{|\lambda|^t}{t}} \\
&= \frac{b^2}{\sigma^2}\{-\log(1-|\lambda|)\} \mbox{ , where $0<|\lambda|<1$.} \\
&= C.b^2 \rightarrow 0 \mbox{ , as $b \rightarrow 0$ , where $C$ is a constant}
\end{array}$


\noindent Thus $MSE(T_b(x))=Var(T_b(x))+Bias(x)^2 \rightarrow 0 \mbox{ , as $n \rightarrow \infty$}$. Hence, we have, $T_b(x) \overset{\mathbb{L}_2}{\longrightarrow} G(x) \mbox{ , $\forall x$ as $n \rightarrow \infty$ }$ which implies the result.
\end{proof}

\textbf{Note 5.1}: For $|x| \leq 1$, $\frac{1}{\sqrt{2\pi}}<1$, $e^{-x^2/2}\leq 1$ implies $|x\phi(x)|<1$. For $|x|>1$, since $|x\phi(x)|=|x|\phi(|x|)$, it can be written as $\frac{1}{\sqrt{2\pi}}\frac{|x|}{e^{|x|^2/2}}=\frac{1}{\sqrt{2\pi}}\frac{1}{\frac{1}{|x|}+\frac{|x|}{2} + \delta} \leq \frac{2}{\sqrt{2\pi}}<1$, where $\delta > 0$.
\subsection{\underline{Proof of Theorem 2.5}} 
\begin{proof}
For $1 \leq i \leq n$, we have,
\begin{eqnarray*}
& & E[(Z_i-E(Z_i))^2(X^{\prime}_i-E(X^{\prime}_i))^2] \\
& = & p E[(Z_i-E(Z_i))^2(X^{\prime}_i-E(X^{\prime}_i))^2 \given B_i=1] \\
& & ~~~~+ (1-p)E[(Z_i-E(Z_i))^2(X^{\prime}_i-E(X^{\prime}_i))^2 \given B_i=0] \\
& = & pE[\frac{1}{n-1} \sum_{j=1,j \neq i}^n{[(X_j-E(X_j))^2(X^{\prime}_i-E(X^{\prime}_i))^2]}] \\
& & ~~~~+(1-p)E[[(X_i+Y_i-E(X_i+Y_i))^2(X^{\prime}_i-E(X^{\prime}_i))^2]] \\
& & \mbox{ [$ \because E[Y_i]=0$ and $E[Z_i]=E[X_j]$ $\forall 1 \leq i,j \leq n$ as shown in Proof of Theorem 2.1 }] \\
& = & p\cdot Var(X^{\prime})\cdot Var(X) + (1-p)\cdot\{ d(X,X^{\prime})+ 2\cdot 0 +Var(Y)\cdot Var(X^{\prime}) \} < \infty
\end{eqnarray*}

Thus, $\hat{Cov}(Z,X^{\prime})=\frac{1}{n}\cdot \sum_{j=1}^n[{Z_j\cdot X^{\prime}_j}]-\bar{Z}\cdot \bar{X^{\prime}}$ exists and hence is a consistent estimator of the true covariance between $Z$ and $X^{\prime}$.
\begin{eqnarray*}
Cov(Z,X^{\prime}) & = & E(ZX^{\prime})-E(Z)\cdot E(X^{\prime}) \\
& = & p E(X)\cdot E(X^{\prime})+ (1-p) E((X+Y)\cdot X^{\prime}) - E(Z)\cdot E(X^{\prime})\\
& = & (1-p) [E(X\cdot X^{\prime})- E(Z)\cdot E(X^{\prime})] \\
& & \mbox{ [$ \because E[Z]=E[X]$ and $E[YX^{\prime}]=E[Y]E[X^{\prime}]=0$]} \\
& = & (1-p). Cov(X,X^{\prime})
\end{eqnarray*}

Thus $\hat{Cov}(Z,X^{\prime})$ is a consistent estimator of $(1-p). Cov(X,X^{\prime})$, or $\frac{1}{1-p}\hat{Cov}(Z,X^{\prime})$ is a consistent estimator of $Cov(X,X^{\prime})$.
Also, we have seen in Theorem 2.1, $Var(Z)$ exists if $Var(X)$ does. Thus $\sqrt{\hat{Var}(Z)} \overset{P}{\longrightarrow} \sqrt{Var(Z)}$ and since $Var(X^{\prime})$ exists, $\sqrt{\hat{Var}(X^{\prime})} \overset{P}{\longrightarrow} \sqrt{Var(X^{\prime})}$
Hence, by property of convergence in probability, the result follows.

\end{proof}


\begin{thebibliography}{99}

\bibitem{JSLP}
J. Steinberg  L Pritzker. \emph{Some experiences with and reactions on data linkage in the united states}, 1967, Bulletin of the International Statistical Institute, 786-808.
\bibitem{RBRB}

R. Bachi  R. Baron. \emph{Confidentiality problems related to data banks}, 1969, Bulletin of the
International Statistical Institute, 43, pp. 225-241.

\bibitem{TD}
T. Dalenius \emph{The invasion of privacy problem and statistics production-an overview}, 1974, Statistisk Tidskrzft, 213-225.A

\bibitem{TDA}
T. Dalenius \emph{Computers and individual privacy some international implications}, 1977a, Bulletin of the International Statistical Institute, 47, pp.203-211.

\bibitem{RHM}
R.H. Mugge \emph{Issues in protecting confidentiality in national health statistics}, 1983, Proceedings of the Section on Survey Research Methods, American Statistical Association, pp.592-594.

\bibitem{SEF}
S.E. Fienberg \emph{Conflict between the needs for access to statistical information and
demands for confidentiality}, 1994, Journal of Official Statistics 10(2) pp.115-132.
\bibitem{TSBM}
S. Trabelsi V. Salzgeber M Bezzi G. Montagnon \emph{Data Disclosure Risk Evaluation}, 2009, IEEE Xplore DOI: \emph{10.1109/CRISIS.2009.5411979}

\bibitem{WAF}
W.A. Fuller \emph{Masking  Procedures for Microdata Disclosure Limitation} Journal of Official Statistics 1993 pp. 383-406

\bibitem{WF}
W.A. Fuller \emph{Measurement Error models}. New York: John Wiley

\bibitem{DR}
T. Dalenius  S. P. Reiss \emph{Data-Swapping: A Technique for Disclosure Control}, 1982, Journal of Statistical Planning and Inference 6, 73-85.

\bibitem{MRA}
R. A. Moore \emph{Controlled data Swapping Techniques for Masking Use Microdata Sets} US Bureau of the Census, Statistical Research Division 1996 \\
Available at {\em http://www.census.gov/srd/www/byyear.html}

\bibitem{RK}
R. Sarathy \& K. Muralidhar \& R. Parsa \emph{Perturbing Non-normal Confidential Attributes: The Copula Approach} Management Science, Vol. 48, No. 12 Dec. 2002 INFORMS pp. 1613-1627 \\
Available at{ \em http://www.jstor.org/stable/822527}

\bibitem{GR}
D. Ghatak \& B. Roy \emph{Estimation of True Quantiles from Quantitative Data Obfuscated with Additive Noise} Journal of Official Statistics \em{In Press}
\end{thebibliography}
\end{document}